\documentclass[twocolumn]{aastex631}

\newcommand{\OVI}{\ion{O}{6}}
\newcommand{\MgII}{\ion{Mg}{2}}
\newcommand{\HA}{H{$\alpha$}}
\newcommand{\CII}{\ion{C}{2}}

\shorttitle{UV Cooling via {\OVI} Emission in the Superwind of {M82} Observed with FUSE}
\shortauthors{Kim et al.}
\graphicspath{./}

\begin{document}

\title{UV Cooling via {\OVI} Emission in the Superwind of {M82} Observed with the Far Ultraviolet Spectroscopic Explorer (FUSE)}

\author[0000-0001-8380-9988]{Jin-Ah Kim}
\affiliation{University of Arizona, Steward Observatory, 933 N. Cherry Ave., Tucson, AZ 85721, USA}

\author[0000-0002-3043-2555]{Haeun Chung}
\affiliation{University of Arizona, Steward Observatory, 933 N. Cherry Ave., Tucson, AZ 85721, USA}

\author[0000-0001-7936-0831]{Carlos J. Vargas}
\affiliation{University of Arizona, Steward Observatory, 933 N. Cherry Ave., Tucson, AZ 85721, USA}

\author[0000-0002-3131-7372]{Erika Hamden}
\affiliation{University of Arizona, Steward Observatory, 933 N. Cherry Ave., Tucson, AZ 85721, USA}



\begin{abstract}

We examined archival Far Ultraviolet Spectroscopic Explorer data to search for far-ultraviolet emission lines in the starburst galaxy {M82}. The observations were made in an outflow region that extends beyond the galactic disk. We found the {\OVI} $\lambda\lambda$ 1032, 1038 emission lines from the galaxy's southern outflow region. The {\OVI} lines suggest that the outflowing warm-hot gas is undergoing radiative cooling. We measured a radial velocity of $\sim$420 km s$^{-1}$ from the {\OVI} lines, which is faster than the velocity seen in {\HA} observations. The {\OVI} $\lambda$1038 emission line seems to be blended with the {\CII} $\lambda$1037 emission line, which has a radial velocity of $\sim$300 km s$^{-1}$, similar to what is observed in {\HA} observations. The outflow medium of M82 appears to be composed of gas in multiple phases with varying temperatures and kinematics. Future spectroscopic observations in high energy regimes covering a wider spatial area are necessary to understand better the properties of the warm-hot gas medium in the outflow.

\end{abstract}

\keywords{Ultraviolet spectroscopy, Galaxies, Circumgalactic medium, Starburst galaxies}


\section{Introduction} \label{sec:intro}

The circumgalactic medium (CGM) is the gas surrounding a galactic disk and filling a galactic halo. The CGM and the interstellar medium are constantly interacting, transforming into each other through mechanisms such as gas accretion and galactic outflows \citep[e.g.,][]{2008MNRAS.387..577O, 2019MNRAS.488.1248H}. This interaction can change the gas available to form stars and metal mixing in galaxies. Understanding the CGM is crucial to comprehending star formation and metal composition in galaxies.

The presence of the CGM was first suggested based on the detection of absorption lines formed while light passes through extended galactic halos \citep[see][for review]{2017ARA&A..55..389T}. Many CGM studies still rely on detecting absorption lines by observing background sources such as quasars or halo stars \citep[e.g.,][]{1969ApJ...156L..63B, 2012ApJ...756L...8G, 2015ApJ...804...79L}. Another method to investigate the CGM is analyzing the continuum or line emission emitted by the CGM directly \citep[e.g.,][]{2013ApJ...762..106A, 2016ApJ...828...49H}. These emission line studies are rare and challenging due to the low surface brightness of the emission. Nevertheless, emission line studies have the advantage of allowing us to explore areas without bright background sources.

The CGM is a complex system consisting of multiple phases of matter, each with varying temperature and velocities \citep[see Figures 4 and 5 for details in][]{2017ARA&A..55..389T}. Various emission and absorption lines emerge from different ionized energy states as illustrated in Figure 6 of \citet{2017ARA&A..55..389T}. Lines associated with warm-hot CGM gas are primarily detected using X-ray or ultraviolet (UV) wavelengths. For example, highly ionized metal lines like {\ion{O}{8}} and {\ion{Mg}{10}} are mostly induced in a hot phase ($\textrm{T}\sim10^{6}~\textrm{K}$), while {\OVI} line is induced in a warm-hot phase ($\textrm{T}\sim10^{5.5}~\textrm{K}$). {\CII} and {\ion{N}{2}} lines occur in a colder phase ($\textrm{T}\sim10^{4.5}~\textrm{K}$). 

{M82} is a starburst galaxy, 3.61 kpc away from us \citep{2018MNRAS.479.4136K}. A strong outflow from the galaxy has been known through observations at multiple wavelengths \citep[e.g.,][]{1963ApJ...137.1005L, 1998ApJ...493..129S, 2009ApJ...697.2030S}. Multiwavelength observations, from X-ray, far-UV (FUV), and visible, through radio wavelengths, show diffuse emission extending beyond the galactic disk. The extended morphologies in these multiple observations are similar to each other and indicate a powerful galactic wind \citep{2015ApJ...814...83L}. Therefore, {M82} is an excellent laboratory to study how a strong galactic wind can affect the CGM.

We analyze Far Ultraviolet Spectroscopic Explorer (FUSE) observational data to examine FUV emission lines around {M82}. In this paper, we present {\OVI} emission lines observed in {M82}'s outflow. In section \ref{sec:data}, we describe the archival data and the data reduction we used. The {\OVI} lines we found are shown in section \ref{sec:result} and discussed in section \ref{sec:discuss}.

\section{Data} \label{sec:data}

\begin{figure}
\centering
\includegraphics[width=\linewidth]{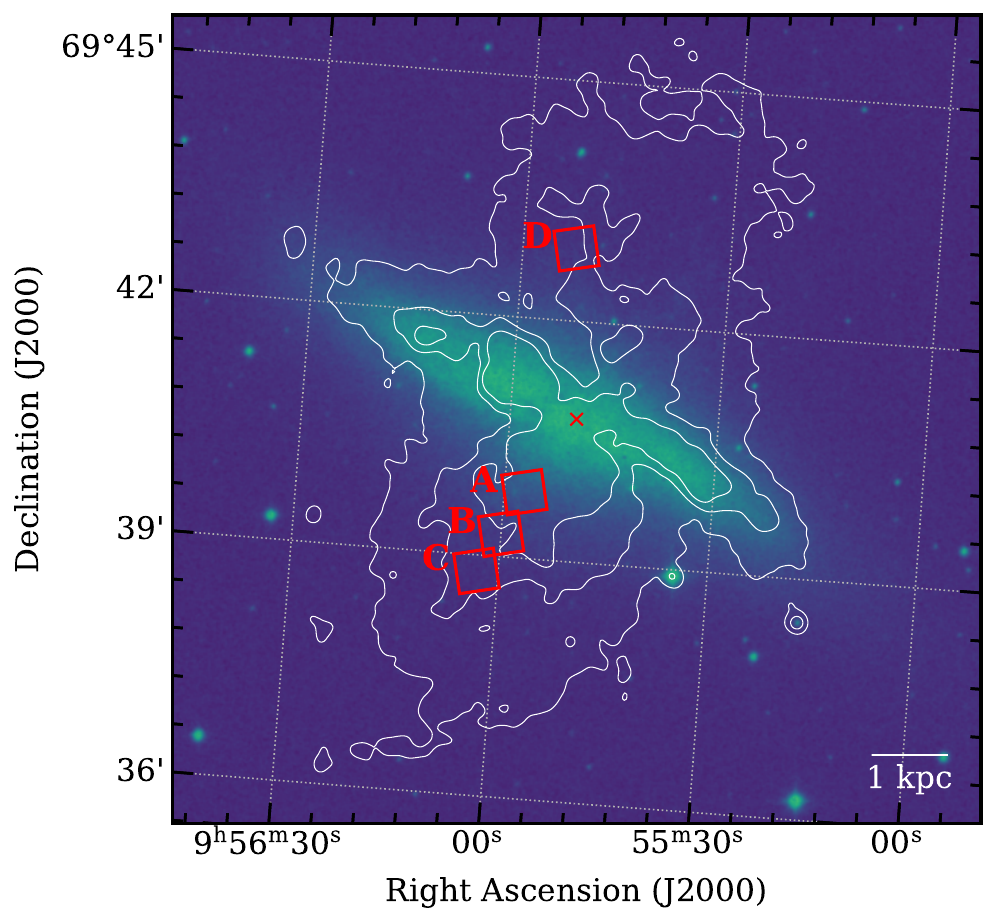}
\caption{Four LWRS observed regions over an optical image around {M82}. The aperture size of LWRS is 30\arcsec × 30\arcsec. A color map shows the DSS POSS-II red optical image. Contour lines are at the GALEX FUV intensity of 0.5, 1, and 2$\times10^{-18}$ erg\;cm$^{-2}$\;s$^{-1}$\;\AA$^{-1}$\;arcsec$^{-2}$.
\label{fig:apertures}}
\end{figure}

FUSE observations were made around the {M82} halo under program C131 in 2002. The observational data used in this study may be accessed from the MAST archive at \dataset[doi:10.17909/y0wa-4b44]{http://dx.doi.org/10.17909/y0wa-4b44}. FUSE observed in four regions near {M82}, shown in Figure \ref{fig:apertures}. Three of them are located on the southern side of the galaxy, and the other region is on the northern side. The regions A, B, and C in Figure \ref{fig:apertures} are 1.2, 1.8, and 2.4 kpc far, respectively, from the galactic center. Region D is around 1.9 kpc offset from the galactic plane.

FUSE observations were made within areas of strong X-ray emission \citep[see Figure 1 in][]{2003ApJ...596L.175H}. The X-ray emission observed in {M82} has a bipolar outflow shape \citep{2004ApJS..151..193S}. The bipolar outflow feature is similar as seen with the GALEX FUV contours in Figure \ref{fig:apertures} and also found by observations at various wavelengths \citep[e.g.,][]{2015ApJ...814...83L}. Note that, although some line emission could be emitted by the warm-hot CGM, the FUV diffuse emission observed by GALEX would be mainly attributed to light originating in the galactic disk and scattering by the extended dust halo \citep{2005ApJ...619L..99H, 2013ApJ...778...79C}.

The FUSE satellite \citep{2000ApJ...538L...1M, 2000ApJ...538L...7S} was launched in 1999 and operated until 2007. FUSE was designed to target FUV wavelengths ranging from around 900 to {1200 \AA}, with a spectral resolving power of roughly 20,000. We specifically analyze data obtained through the low-resolution (LWRS) aperture (30\arcsec × 30\arcsec), the largest aperture of FUSE. Our analysis is based on data collected in LiF1A and LiF2B channels, which cover {\OVI} emission lines at 1032 and {1038 \AA}. The LiF1A channel covers from 987 to {1082 \AA}, and the LiF2B channel covers from 979 to {1075 \AA}. Although the SiC1A and SiC2B channels also cover these wavelengths, we did not incorporate the data from the SiC channels in our study. The presence of solar-scattered {\OVI} emission lines in the SiC channels can interfere with data analysis. The LiF channels are located on the instrument's side, facing away from the Sun to minimize emission lines from solar-scattered light. The effective area is also far lower in the SiC1A and SiC2B channels than in the LiF1A and LiF2B channels, resulting in a lower signal-to-noise ratio (S/N) in the SiC1A and SiC2B channels. Due to these two effects, the LiF1A and LiF2B channels are better suited for studying {\OVI} lines than the SiC channels.

We processed the FUSE observational data through CalFUSE version 3.2.3, a data reduction pipeline for the FUSE data. CalFUSE is described in \citet{2007PASP..119..527D}. We reduced each RAW data file and generated intermediate data files (IDF files). We then combined and calibrated the multiple IDF files using the data analysis tools within the CalFUSE pipeline package, specifically \texttt{idf\_combine} and \texttt{cf\_extract\_spectra}. We used CalFUSE's default parameters but set the initial wavelength for LiF Channels (LIF\_W0) to 987.1 Å. This was done to ensure that the spectra from the LiF1A and LiF2B channels started at the same wavelength. We rebinned the calibrated data using \texttt{cf\_arith} to have a wavelength bin size of {0.104 \AA}~(16 detector pixels). The CalFUSE calibrated files provide fluxes, flux errors, raw photon counts, and estimated background counts as a heliocentric wavelength spectrum. The reduced data products are available in \dataset[doi:10.5281/zenodo.10680806]{http://dx.doi.org/10.5281/zenodo.10680806}.

Our main interest in this study is {\OVI} $\lambda$1031.9 ($1s^2~2s~^2S_{1/2}-1s^2~2p~^2P_{3/2}$) and $\lambda$1037.6 ($1s^2~2s~^2S_{1/2}-1s^2~2p~^2P_{1/2}$) emission lines, which are primarily emitted from the warm-hot CGM at $\textrm{T}\sim10^{5.5}~\textrm{K}$. \citet{2003ApJ...596L.175H} also studied FUSE observations for {M82} but found no detectable {\OVI} $\lambda$1032 emission lines. They used data reduced by an older version of CalFUSE (2.0.5). Several parameters and functions have been updated since version 2.0.5. \citep[see][for details]{2007PASP..119..527D}. One noticeable change is the range of acceptable pulse height values. The pulse height used in \citet{2003ApJ...596L.175H} is between 4 and 14. The default pulse height limits in CalFUSE version 3.2.3. are from 2 to 25 in the LiF1A channel and 3 to 24 in the LiF2B channel. \citet{2007PASP..119..527D} point out that narrower pulse height limits than the default values can cause significant flux losses. \citet{2021ApJ...916....7C} also show that the updated pulse height limit helps detect emission lines invisible with the old pulse height limit. Here we revisit the FUSE observational data using the up-to-date CalFUSE pipeline and see if any spectral lines have been missed.

\section{Results} \label{sec:result}

\begin{figure*}[t]
\centering
\includegraphics[width=\linewidth]{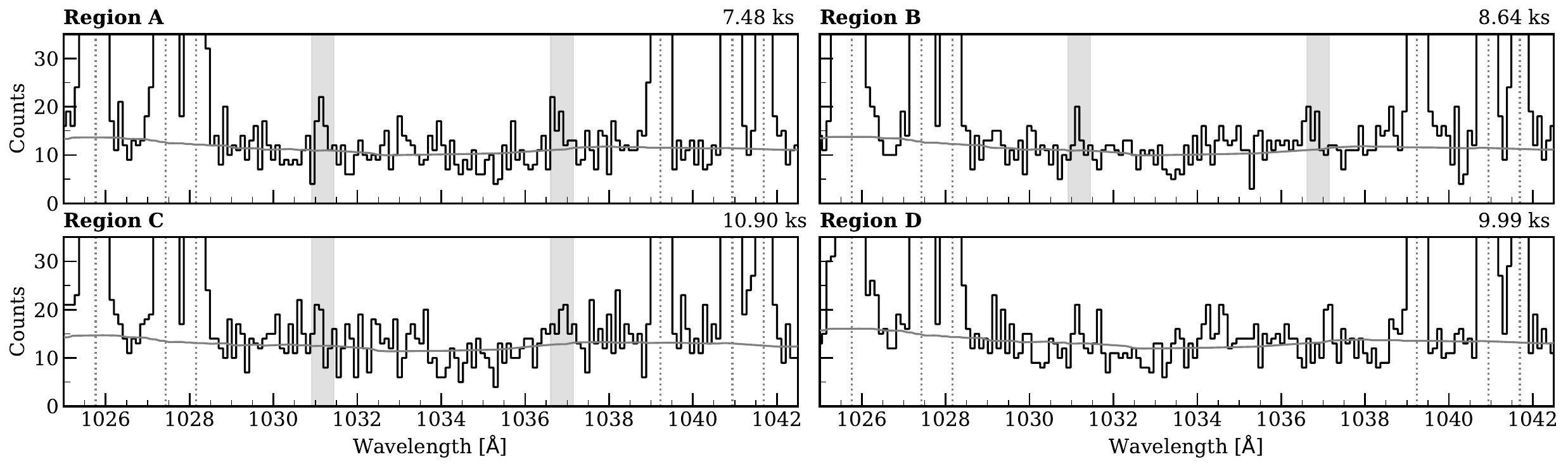}
\caption{FUSE spectra from the LiF2B channel. Gray shaded regions indicate potential locations of the {\OVI} $\lambda\lambda$ 1032, 1038 emission lines. The width of the shades is {0.53 \AA}. A width of {0.53 \AA}~covers $\sim$90\% of the emission flux when an intrinsic line FWHM is $\sim$81 km s$^{-1}$. Horizontal solid gray lines depict background values. Vertical dotted lines are positioned at the geocoronal emission lines. On the top-right corner of each spectrum, the number is a total exposure time, including both day and night exposures.
\label{fig:spectrum}}
\end{figure*}

Figure \ref{fig:spectrum} displays the individual FUSE spectra of four regions in M82. We plotted in the spectra the dead-time corrected photon count values (referred to as WEIGHTS in CalFUSE) from CalFUSE calibrated files created using the LiF2B channel data. From the data on the southern side (Regions A, B, and C), we found potential emission peaks at around 1031 and {1037 \AA}. The two peaks are expected to correspond to blue-shifted {\OVI} $\lambda\lambda$ 1032, 1038 emission lines. The locations of two potential lines are marked as gray shades, and the lines are shifted as much as the V$_{\tiny \textrm{LOS}}$ value from Table \ref{tab:fitparamtwoOVI}. A S/N estimated in each shaded region is less than 3, while the longer wavelength line in Regions A and C has a S/N higher than 2.5. The average of S/Ns is approximately 2. Despite the low S/N, the line feature consistently seen in three regions makes the presence of the {\OVI} emission lines promising. The spectra from the LiF1A channel data have similar peaks, but their S/Ns are lower than those in the LiF2B spectra.

The spectrum of the northern side (Region D) does not have apparent emission line features. In Region D of Figure \ref{fig:spectrum}, S/Ns at the shaded wavelength regimes are less than 1. In particular, the LiF1A spectrum at Region D is dominated by background noise. This is likely due to the fact that the outflow on the northern side is positioned behind the northwest side of the galaxy disk, which is closer to us \citep{1963ApJ...137.1005L}. As the emission from the outflow on the northern side should pass through the galactic disk, it is subject to greater extinction than the emission on the southern side. As a result, the emission lines on the northern side may be obscured and may not be visible.

To enhance the S/N in a resultant spectrum, we combined the spectra from three regions A, B, and C using \texttt{idf\_combine}. As all the observations were made between 2002-02-02 and 2002-02-03, there was no significant change in effective area of the telescope during this short period. Outflow velocity measured in {M82} with optical observations changes slowly beyond a distance of around 30\arcsec~from the center \citep{1998ApJ...493..129S}. Because of the opposite outflow directions on the northern and southern sides of the galaxy, we combined only the data from the southern side.

\begin{figure}
\centering
\includegraphics[width=0.95\linewidth]{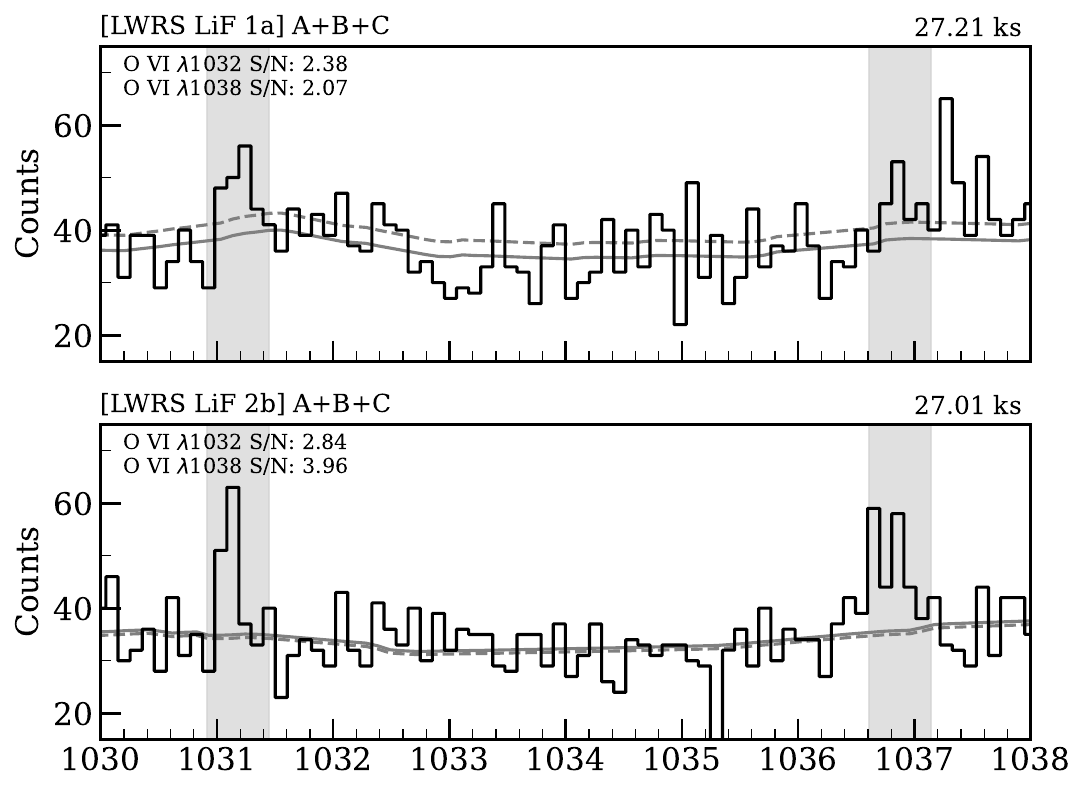}
\caption{FUSE spectra combined from the observations on the southern side. The top and bottom panels are the spectrum from the LiF 1A and 2B channels, respectively. Gray shades are the same as in Figure \ref{fig:spectrum}. Horizontal gray lines are background values. The dashed lines are the background values calculated through CalFUSE, and the solid lines are the adjusted values given our MCMC fitting result.
\label{fig:comb_spectrum}}
\end{figure}

Figure \ref{fig:comb_spectrum} shows the count spectrum combined from observational data for Regions A, B, and C, separately in the LiF 1A and 2B channels. Both channels show a clear emission peak near {1031.2 \AA}. The other peak is located near {1037.9 \AA}, although the peak locations in LiF 1A and 2B channels are slightly offset. Because of the low S/N, the line position and flux estimated in each combined spectrum may not be precise. For more accurate results, we incorporate the spectra from both channels to estimate the physical properties of the emission lines.

\begin{figure}
\centering
\includegraphics[width=0.95\linewidth]{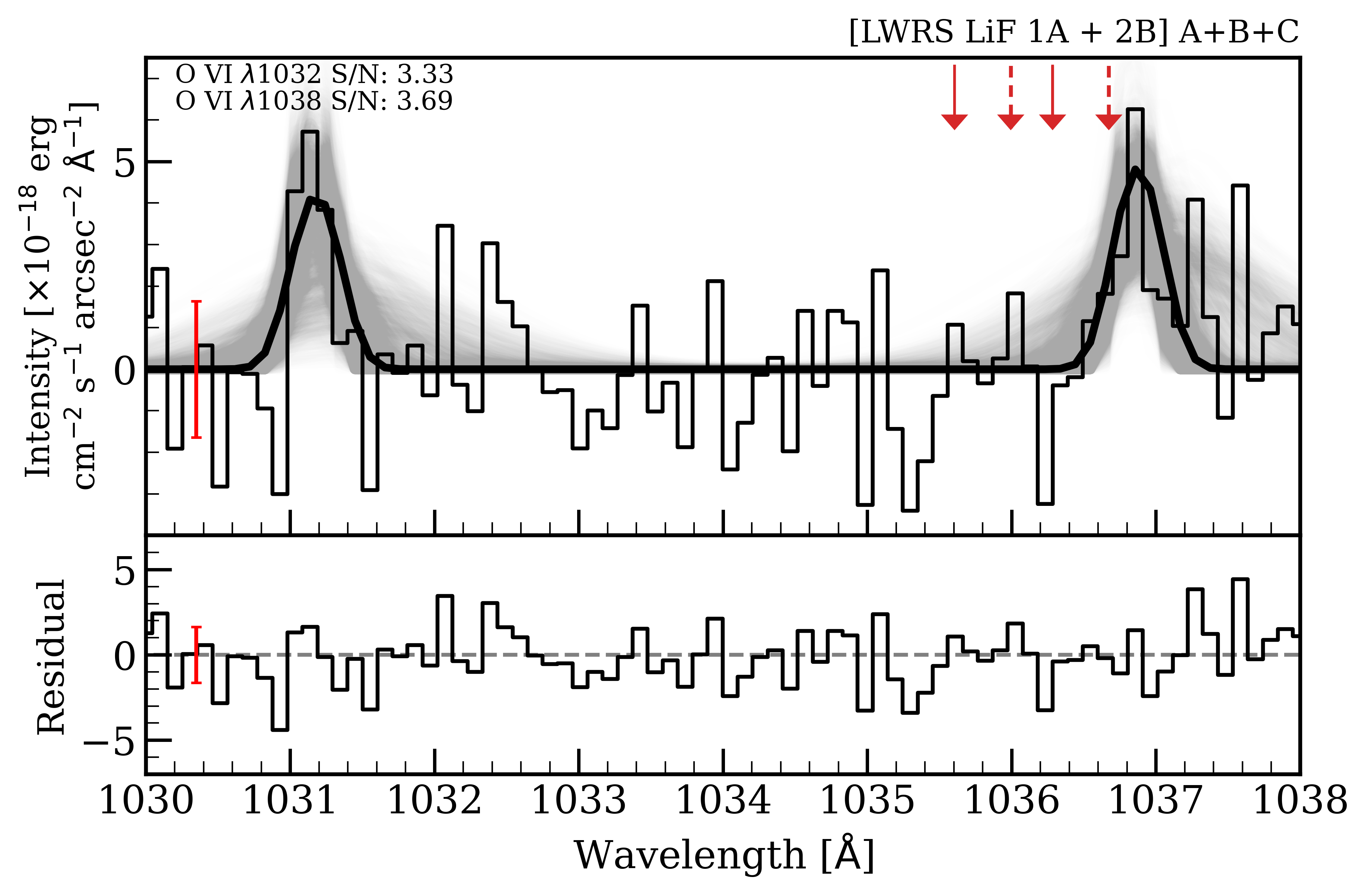}
\caption{Model spectra plotted over the observations. The observation spectrum is combined from the LiF 1A and 2B spectra. The thick black spectrum is modeled with the median values in Table \ref{tab:fitparamtwoOVI}. Gray lines depict sample spectra drawn from the MCMC chains. Red arrows in the top panel indicate the possible locations of {\CII} $\lambda\lambda$ 1036, 1037. Solid red arrows are with the radial velocity in Table \ref{tab:fitparamtwoOVI}. Dashed red arrows are shifted with a radial velocity of -100 km s$^{-1}$, taken from {\HA} observations \citep{1998ApJ...493..129S}. The bottom panel displays the deviation between the observation and the median model spectrum. Red error bars represent a intensity uncertainty in the spectrum's wavelength range.
\label{fig:model_spectrum}}
\end{figure}

We used a Markov chain Monte Carlo (MCMC) approach to derive properties of {\OVI} $\lambda\lambda$ 1032, 1038 emission lines. We created an intrinsic model spectrum containing the {\OVI} doublet lines, which are parameterized with a heliocentric radial velocity (V$_{\tiny \textrm{LOS}}$), a full width at half maximum (FWHM) of lines, and individual line intensities (I$_{\textrm{\footnotesize{\OVI}}, \lambda1032}$ and I$_{\textrm{\footnotesize {\OVI}}, \lambda1038}$). The model spectrum was convolved with a Gaussian profile ({0.021 \AA}) corresponding to the FUSE line spread function and a top-hat function with a {0.362 \AA}~width accounting for the filling effect on the LWRS aperture. We converted the values in the model spectrum from flux to photon counts and added background values. The added background values are derived from multiplying the background count spectrum, provided by CalFUSE, and a scaling factor, which is one of fitting parameters. The scaling factors correct the overestimated background shown in the LiF1A spectrum of Figure \ref{fig:comb_spectrum} and account for the continuum in the wavelength regime. The adjusted background values according to the fitting result are shown as solid horizontal lines in Figure \ref{fig:model_spectrum}. Lastly, using a continuous Poisson probability function, $P(N_{obs}) \propto (N_{model})^{N_{obs}}~exp(-N_{model}) / \Gamma(N_{obs}+1) $, we estimated the likelihood of the observational data given the model spectrum separately for the LiF 1A and 2B channels. We performed MCMC fitting, assuming the product of individual likelihoods in both channels as the likelihood of observational data given model spectra. We estimated the likelihood separately in the LiF 1A and 2B channels because of the different background values and the different flux conversion factors from different effective areas. Table \ref{tab:fitparamtwoOVI} presents the MCMC results including the median value and the range of the 68\% confidence level of parameter probability distributions.

Figure \ref{fig:model_spectrum} compares the model spectra and the observational spectrum. The model spectrum plotted in black was created using the median values for each model parameter in Table \ref{tab:fitparamtwoOVI}. The observational spectrum is generated from the flux spectra in the CalFUSE outputs of the LiF 1A and 2B channels. We combined the flux spectra from the two channels by subtracting the flux, corresponding to the background adjusting values from the MCMC result, and weighting by the inverse square of the flux errors. The deviation between the observational spectrum and the median model spectrum is shown in a lower panel of Figure \ref{fig:model_spectrum}.

Based on our fitting result, it appears that I$_{\textrm{\footnotesize {\OVI}}, \lambda1032}$ is lower than I$_{\textrm{\footnotesize {\OVI}}, \lambda1038}$. This is unexpected, because these lines are resonance doublets, therefore {\OVI} ${\lambda1032}$ line should be two times brighter than {\OVI} ${\lambda1038}$ line. One possible explanation for this unexpected result is that another emission line may be blended with the {\OVI} $\lambda$ 1038 line, leading to in an incorrect estimation of I$_{\textrm{\footnotesize {\OVI}}, \lambda1038}$. We will discuss this further in Section \ref{subsec:OVI1038}.

Note that the line FWHM may not accurately represent properties of the medium emitting the {\OVI} lines. The top-hat function with a {0.362 \AA}~width is estimated when emission sources extensively fill an observing aperture. If the emission is locally concentrated in the aperture, the intrinsic line FWHM may be underestimated in our method. Also, since the spectrum is combined from three different regions, the velocity variation in these regions may appear as a larger line FWHM.

\begin{deluxetable*}{cccc}
\tablecaption{MCMC results including {\OVI} $\lambda\lambda$1032, 1038 emission lines. \label{tab:fitparamtwoOVI}}
\tablehead{
V$_{\tiny \textrm{LOS}}$ - V$_{\tiny \textrm{sys,{M82}}}$ & FWHM & I$_{\textrm{\footnotesize {\OVI}}, \lambda1032}$ & I$_{\textrm{\footnotesize {\OVI}}, \lambda1038}$ \\
{[km s$^{-1}$]} & {[km s$^{-1}$]} & {[erg cm$^{-2}$\;s$^{-1}$\;arcsec$^{-2}$]} & {[erg cm$^{-2}$\;s$^{-1}$\;arcsec$^{-2}$]}
}
\startdata
& \\ [-1.7ex]
-415.67 $^{+42.69}_{-12.65}$ & 81.42 $^{+216.17}_{-53.13}$ & 1.78 $^{+0.66}_{-0.52}$ $\times10^{-18}$ & 2.05 $^{+1.64}_{-0.66}$ $\times10^{-18}$\\ [1.3ex]
\enddata
\tablecomments{Median values and 68\% confidence ranges from parameter probability distributions. The velocities are heliocentric. The velocity of {M82} (V$_{\textrm{\tiny sys, {M82}}}$) is 203 km s$^{-1}$ \citep{1995yCat.7155....0D}. Intensities are uncorrected for extinction.}
\end{deluxetable*}

\begin{figure}
\centering
\includegraphics[width=0.7\linewidth]{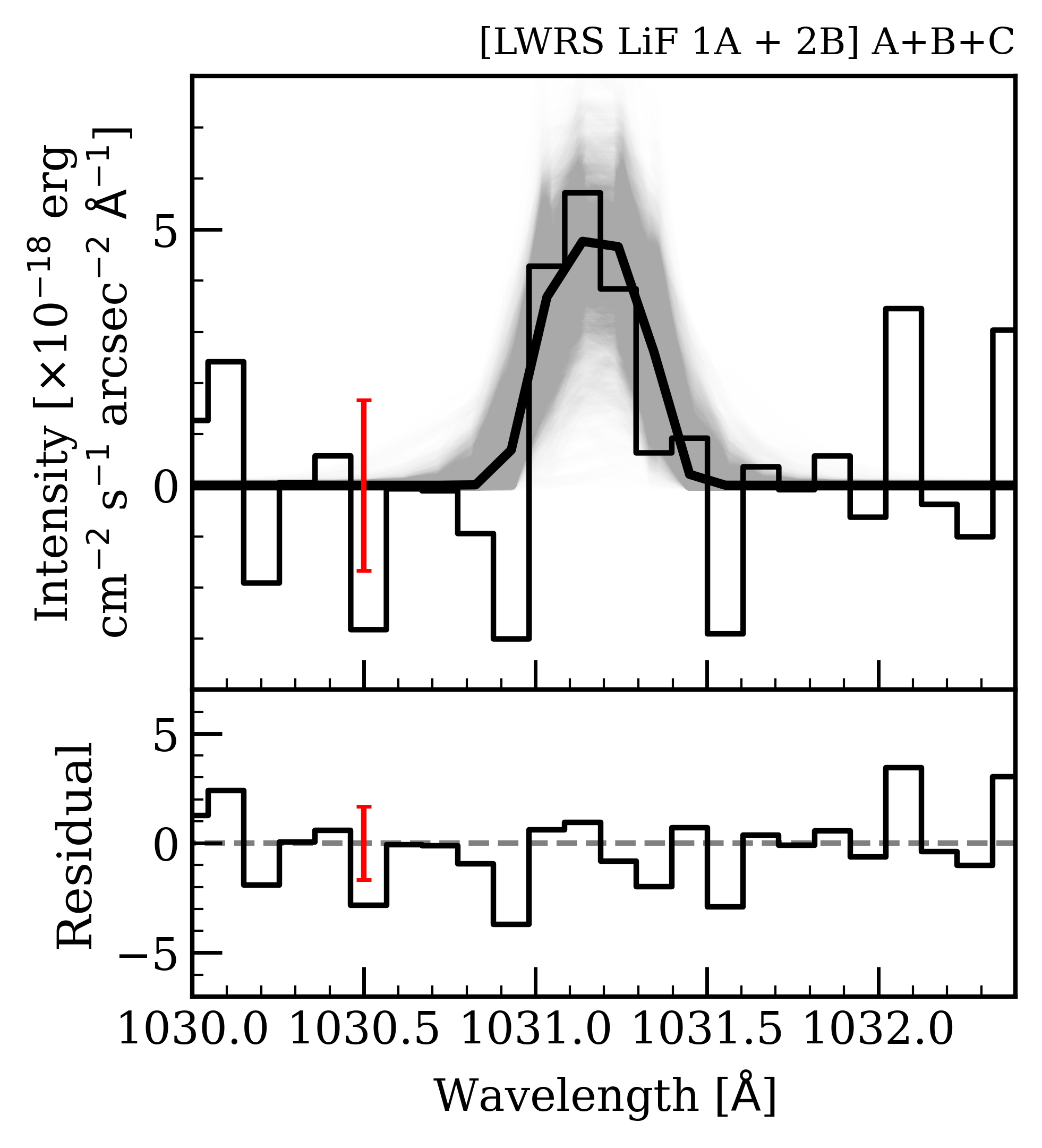}
\caption{The same as Figure \ref{fig:model_spectrum}, but only including fitting for {\OVI} $\lambda$1032 emission line. The model parameters can be found in Table \ref{tab:fitparamoneOVI}.
\label{fig:singlemodel_spectrum}}
\end{figure}

Additionally, we repeated the MCMC fitting process, focusing solely on the {\OVI} $\lambda$1032 emission line. The results in Table \ref{tab:fitparamtwoOVI} are estimated assuming that the FWHM is consistent for both {\OVI} $\lambda$1032 and $\lambda$1038 lines. However, if the {\OVI} ${\lambda1038}$ line is blended with other lines, as mentioned previously, the line FWHM and intensity observed at {\OVI} $\lambda$1038 would not be exclusively attributed to the {\OVI} ${\lambda1038}$ emission line. Consequently, it is unnecessary for the {\OVI} $\lambda$1032 and $\lambda$1038 lines to have the same line FWHM. It is also challenging to determine the properties of the {\OVI} ${\lambda1038}$ emission line. The fitting results only of {\OVI} $\lambda$1032 emission line are presented in Figure \ref{fig:singlemodel_spectrum} and Table \ref{tab:fitparamoneOVI}. The radial velocity and intensity of the line remain unchanged from the previous fitting of the {\OVI} $\lambda\lambda$1032, 1038 doublet. However, the line FWHM is noticeably narrower when derived from the single line fitting. This is also a potential evidence that another emission line may overlap with the {\OVI} $\lambda$1038 emission line, causing it to appear broader in the spectrum.

\begin{deluxetable*}{ccc|c}
\tablecaption{MCMC results only for {\OVI} $\lambda$1032 emission line and an upper limit for the {\CII} $\lambda$1037 emission line.
\label{tab:fitparamoneOVI}}
\tablewidth{0pt}
\tablehead{
\multicolumn{3}{c|}{{\OVI} $\lambda$1032} & \multicolumn{1}{c}{{\CII} $\lambda$1037}
}
\startdata
V$_{\tiny \textrm{LOS}}$ - V$_{\tiny \textrm{sys,{M82}}}$ & FWHM & I$_{\textrm{\footnotesize {\OVI}}, \lambda1032}$ & I$_{\textrm{\footnotesize {\CII}}, \lambda1037}$  \\
{[km s$^{-1}$]} & {[km s$^{-1}$]} & {[erg cm$^{-2}$\;s$^{-1}$\;arcsec$^{-2}$]} & {[erg cm$^{-2}$\;s$^{-1}$\;arcsec$^{-2}$]}\\
 \hline
 & & & \\ [-1.7ex]
-418.16 $^{+0.10}_{-0.10}$ & 30.25 $^{+33.91}_{-19.89}$  & 1.73 $^{+0.46}_{-0.45}$ $\times 10^{-18}$ & {$<$ 1.20 $\times 10^{-18}$} \\ [1.3ex]
\enddata
\tablecomments{The same as Table \ref{tab:fitparamtwoOVI} but the fitting results only for {\OVI} $\lambda$1032 emission line. I$_{\textrm{\footnotesize {\CII}}, \lambda1037}$ is estimated as a 2-$\sigma$ upper limit based on the intensity uncertainty and a wavelength range of {0.37 \AA}.}
\end{deluxetable*}

\section{Discussion}\label{sec:discuss}

\subsection{Comparison with other wavelength observations}\label{subsec:otherobs}

The {\OVI} emission lines we observed are most likely emitted from the outflow of {M82}. Based on our observations with the {\OVI} lines, the warm-hot outflow velocity on the southern side of {M82} is approximately -420 km s$^{-1}$. While it is conceivable that high-velocity clouds within the Milky Way may produce these emission lines, the lack of detected emission lines on the northern region leads us to consider that {M82} is the more likely source. Also, according to \citet{2018MNRAS.474..289W}, the radial velocity in the local standard of rest of high-velocity clouds around the line of sight toward {M82} is around 180 km s$^{-1}$.

The spectral lines in the outflow on the southern side of {M82} have been found with multiple observations. \citet{1998ApJ...493..129S} discovered {\HA} lines at a velocity of about -30 and -300 km s$^{-1}$ and another optical emission lines, \ion{N}{2} and \ion{O}{3}, with a similar velocity trend. CO lines observed by \citet{2013PASJ...65...66S} are shifted by around -30 and -100 km s$^{-1}$ relative to the galactic systemic velocity. Note that the outflow velocities mentioned here are line-of-sight velocities, so they are lower than the actual outflow velocity due to the inclination effect.

Our result shows that the radial velocity of {\OVI} emission lines is higher than that of optical and CO emission lines. The {\OVI} emission lines are primarily emitted from a hotter medium ($\textrm{T}\sim10^{5.5}~\textrm{K}$) than optical and CO emission lines are. Therefore, our finding indicates that a hotter phase medium is being expelled faster than a colder medium. Numerical simulations also demonstrated that different phases could have varying velocities. For example, models such as the {M82}'s outflow find that hotter gas moves faster than colder gas \citep{2020ApJ...895...43S}. Bow shock simulations, although simulating a supersonic motion of galaxies, also exhibit reduced velocities in regions with lower temperatures \citep{2011A&A...531A..13S}. Another galactic outflow simulation proposed that {\HA} emission may come from gas cloudlets, broken away from the gas cloud being expelled from a galactic disk, while {\OVI} emission occurs around the gas cloudlets \citep{2008ApJ...674..157C, 2009ApJ...703..330C}. To better understand a multiphase gas outflow's complex structure, more observations and numerical analysis would be required.

Another noticeable difference between our findings and optical or radio line observations is line splitting; that is, each emission line is shifted by two different velocities. The line splitting seen in optical and CO observations is explained as emission from the surface of a biconical outflow \citep[see Figure 8 in][]{2013PASJ...65...66S}. On the other hand, the line splitting is not seen in our FUSE spectrum. However, our observation of a single {\OVI} emission line does not contradict the bi-conical model. The gas medium filling the cones could obstruct the detection of another {\OVI} line emitted from the rear side of the cone due to the high extinction in UV wavelengths. With assumptions that R$\rm_{V}$ of 3.1 and {\HA}/H{\footnotesize $\beta$} of 4 \citep{1990ApJS...74..833H} and using equations in \citet{1989ApJ...345..245C} and \citet{2012MNRAS.421..486X}, the optical depth of the total medium in {M82}'s outflow is approximately 4 at {1035 \AA}. If the front-side emission is radiated in front of the outflow medium and the rear-side emission passes through the total medium, the ratio between the front- and rear-side emission flux is roughly 50. With the low S/N of the observation data we used, it is likely that the fainter emission peaks would be undetected. It is also possible that the front and rear sides of the region emitting {\OVI} are not clearly distinct, unlike the colder phase medium. To better understand the structural difference between the region emitting {\OVI} and the colder gas medium, we would require additional observations that are spatially resolved across a wide field at multiple wavelengths.

\subsection{Flux at \textrm{{\OVI}} \texorpdfstring{$\lambda$}{lambda}1038}\label{subsec:OVI1038}

In the results in Table \ref{tab:fitparamtwoOVI}, the line at {\OVI} $\lambda$1038 is brighter than that at {\OVI} $\lambda$1032. The emission line at {\OVI} $\lambda$1038 also looks broader than that at {\OVI} $\lambda$1032 in Figure \ref{fig:model_spectrum}. Theoretically, when electrons are mainly excited by collisions, the flux ratio of the doublet lines is 2:1, with {\OVI} $\lambda$1032 being brighter. The variation of a flux ratio of doublet lines, similar to {\OVI} doublet lines, also has been observed in star-forming galaxies \citep[e.g.,][]{2018ApJ...855...96H, 2023ApJ...943...94X}. However, the flux ratio lower than 1 is uncommon.

The observed flux ratio of doublet lines can vary depending on optical depths due to self-absorption \citep[e.g.,][]{2001ApJ...560..730S, 2020MNRAS.498.2554C}. \citet{2023arXiv231017908S} performs simulations for {\MgII} doublet lines, similar to {\OVI} doublet lines, that are impacted by self-absorption and scattering effects. However, the flux ratio of the simulated doublets in emission spectra is not found to be less than 1.2. In cases with a very high expansion velocity and continuum sources, the flux ratio of the simulations goes lower than 1. Nevertheless, our observations do not show noticeably enhanced continuum levels to support the cases. Additionally, the contribution of ions excited by photoionization could decrease the flux ratio close to 1 \citep{2001ApJ...560..730S}, but not less than 1. Therefore, it is difficult to explain our observed flux ratio, less than 1, with a simple model.

Another plausible explanation for brighter flux at our {\OVI} $\lambda$1038 is that the line is overlapped with another emission lines, most likley {\CII} $\lambda$1036.3 ($2s~2p^2~^2S_{1/2}-2s^2~2p~^2P_{1/2}$) and {\CII} $\lambda$1037.0 ($2s~2p^2~^2S_{1/2}-2s^2~2p~^2P_{3/2}$). {\CII} $\lambda$1037 emission line has been observed in the Galactic halo \citep{2001ApJ...560..730S, 2002ApJ...569..758S}. \citet{2006ApJ...647..328D} also mention that the {\OVI} $\lambda$1038 line is often possible to be blended with the {\CII} $\lambda$1037 line. In Figure \ref{fig:model_spectrum}, we illustrate possible locations of {\CII} emission lines. The {\CII} lines, which are shifted in the same way as the observed {\OVI} lines (solid arrows), are positioned apart from the {\OVI} lines. However, the plausible {\CII} $\lambda$1037 line shifted same as the {\HA} velocity (dashed arrows), V$_{\textrm{\tiny LOS}} \sim -100 \textrm{ km s}^{-1}$, is close to the {\OVI} $\lambda$1038 line. {\CII} emission lines are effectively emitted from the regions of T$\sim10^{4.5}$ K \citep[Figure 6 in][]{2017ARA&A..55..389T} which is close to the temperature for emitting {\HA} emission. Therefore, {\CII} emitters may have different velocities than {\OVI} emitters but similar to {\HA} emitters. Other UV emission lines, which are primarily emitted at a similar temperature with {\CII} lines, like \ion{C}{3} $\lambda$977.02 and \ion{N}{3} $\lambda$989.79 lines, are hard to be detected due to their proximity to geocoronal emission lines.

\subsection{{\OVI} emitters}\label{subsec:OVIemitter}

In the FUSE spectrum, we are unable to detect {\CII} $\lambda$1037 emission lines at the wavelength where the emission were shifted by the same amount as the observed {\OVI} lines. Instead, we estimate an upper intensity limit in the regime where the {\CII} emission line at the same radial velocity as the observed {\OVI} lines would be expected. This limit has been calculated based on intensity uncertainties, within a wavelength range of {0.37 \AA}. The width of the range covers more than 90\% of the total flux of a line with a FWHM of 30 km s$^{-1}$. The estimated upper limit is noted in Table \ref{tab:fitparamoneOVI}. Accordingly, a lower limit of the brightness ratio between {\OVI} $\lambda$1032 and {\CII} $\lambda$1037 lines is 1.4. We have compared this intensity ratio with various models to speculate about the physical properties of the {\OVI} emitting regions.

By comparing with models in \citet{1993ApJ...407...83S}, we estimate that the temperature is higher than $10^{5.3}$ K if the {\OVI} lines are emitted by a turbulent mixing layer. We also compare the brightness ratio with the shock models in the Mexican Million Models database (shock part; 3MdB$^s$), specifically the model results from \citet{2008ApJS..178...20A} and \citet{2019RMxAA..55..377A}. As per the model, a shock velocity lower than 100 km/s is insufficient to produce the lower limit of the brightness ratio. A shock velocity higher than approximately 130 km s$^{-1}$ is more likely to account for the brightness ratio. On the other hand, infrared observations suggested that the warm H$_2$ outflow may have lower velocity shock, which is lower than 40 km s$^{-1}$ \citep{2015MNRAS.451.2640B}. As shown in the different outflowing velocities between {\OVI} and molecular observations, the {\OVI} emitters and molecular outflow region are likely to be separated.

\section{Conclusion}

We explore the outflow in {M82} using FUSE observations. We found {\OVI} $\lambda\lambda$1032, 1038 emission lines in the southern side of {M82} using FUSE observations. These UV emission lines indicate that the outflowing medium in the warm gas phase is undergoing radiative cooling. The {\OVI} emission lines are shifted by higher radial velocity than {\HA} or CO observations. The higher flux at {\OVI} $\lambda$1038 than one at {\OVI} $\lambda$1032 could be explained by the presence of {\CII} emission lines at different line-of-sight velocity from {\OVI} lines. It is likely that the outflowing medium in the outflow of {M82} has a complex velocity phase in different temperature phases. Therefore, in addition to the previous studies mostly focused on the outflow medium in the colder phase observed at visible or longer wavelengths, more advanced observations at higher energy regimes are required to definitively determine the behavior of outflowing gas.

\begin{acknowledgments}

This work is supported by NASA under award No. 80NSSC23K1129. Some of the data presented in this paper were obtained from the Mikulski Archive for Space Telescopes (MAST) at the Space Telescope Science Institute. The specific observations analyzed can be accessed via \dataset[doi:10.17909/y0wa-4b44]{http://dx.doi.org/10.17909/y0wa-4b44}. STScI is operated by the Association of Universities for Research in Astronomy, Inc., under NASA contract NAS5–26555. Support to MAST for these data is provided by the NASA Office of Space Science via grant NAG5–7584 and by other grants and contracts. This work is based on observations made with the NASA-CNES-CSA Far Ultraviolet Spectroscopic Explorer. FUSE is operated for NASA by the Johns Hopkins University under NASA contract NAS5-32985. This work includes observations made with the NASA Galaxy Evolution Explorer. GALEX is operated for NASA by the California Institute of Technology under NASA contract NAS5-98034. The Second Palomar Observatory Sky Survey (POSS-II) was made by the California Institute of Technology with funds from the National Science Foundation, the National Aeronautics and Space Administration, the National Geographic Society, the Sloan Foundation, the Samuel Oschin Foundation, and the Eastman Kodak Corporation. 

\end{acknowledgments}

%

\vspace{5mm}
\facilities{MAST (FUSE, GALEX)}


\software{Astropy \citep{2013A&A...558A..33A,2018AJ....156..123A, 2022ApJ...935..167A}, \texttt{emcee} \citep{2013PASP..125..306F}, Matplotlib \citep{Hunter:2007}, NumPy \citep{harris2020array}, regions \citep{2022zndo...7259631B}, SciPy \citep{2020SciPy-NMeth}
          }
          

\bibliography{biblio}{}
\bibliographystyle{aasjournal}




\end{document}